%% file: workshop.tex
\documentclass[sigconf]{acmart}

\usepackage{subfigure}
\usepackage{multirow}

\AtBeginDocument{%
  \providecommand\BibTeX{{%
    \normalfont B\kern-0.5em{\scshape i\kern-0.25em b}\kern-0.8em\TeX}}}

\setcopyright{acmcopyright}
\copyrightyear{2022}
\acmYear{2022}
\acmDOI{XXXXXXX.XXXXXXX}

\acmConference[CIKM '22]{31st ACM International Conference on Information and Knowledge Management}{October 17-22,
  2022}{Atlanta Georgia, USA}
\acmPrice{15.00}
\acmISBN{978-1-4503-XXXX-X/18/06}

\begin{document}

\title{Spatiotemporal-Enhanced Network for Click-Through Rate Prediction in Location-based Services}

\author{Shaochuan Lin,
        Yicong Yu,
        Xiyu Ji,
        Taotao Zhou,
        Hengxu He\textsuperscript{*},
        Zisen Sang}
\author{
        Jia Jia\textsuperscript{*},
        Guodong Cao,
        Ning Hu}
\affiliation{%
  \institution{Alibaba Group}
  \city{Hangzhou\&Shanghai\&Beijiing}
  \country{China}
}
\email{{lin.lsc,
        yicongyu.yyc,
        jixiyu.jxy,
        hengxu.hhx,
        jj229618,
        guodong.cao,
        huning.hu}@alibaba-inc.com}
\email{taotao.zhou@lazada.com,zisen.szs@koubei.com}
\thanks{ * Correspoding author}

\renewcommand{\shortauthors}{Lin and Yu, et al.}

\begin{abstract}
 In Location-Based Services(LBS), user behavior naturally has a strong dependence on the spatiotemporal information, $i.e.$, in different geographical locations and at different times, user click behavior will change significantly. Appropriate spatiotemporal enhancement modeling of user click behavior and large-scale sparse attributes is key to building an LBS model. Although most of existing methods have been proved to be effective, they are difficult to apply to takeaway scenarios due to insufficient modeling of spatiotemporal information. In this paper, we address this challenge by seeking to explicitly model the timing and locations of interactions and proposing a \textbf{S}patio\textbf{t}emporal-\textbf{E}nhanced \textbf{N}etwork, namely StEN. In particular, StEN applies a Spatiotemporal Profile Activation module to capture common spatiotemporal preference through  attribute features. A Spatiotemporal Preference Activation is further applied to model the personalized spatiotemporal preference embodied by behaviors in detail. Moreover, a Spatiotemporal-aware Target Attention mechanism is adopted to generate different parameters for target attention at different locations and times, thereby improving the personalized spatiotemporal awareness of the model. Comprehensive experiments are conducted on three large-scale industrial datasets, and the results demonstrate the state-of-the-art performance of our methods. In addition, we have also released an industrial dataset for takeaway industry to make up for the lack of public datasets in this community.

\end{abstract}

\ccsdesc[500]{Information systems~Location based services}
\ccsdesc[500]{Information systems~Recommender systems}

\keywords{spatiotemporal systems, click-through rate prediction, location-based services}

\maketitle

\section{Introduction}
Location-Based Services (LBS) are mobile services that provide the user with current location-relevant content on smartphones or other services. Among them, takeaway service is the most popular and convenient commercial service. Like other LBS, it also requires timely delivery, which results in a strong dependence on time and geographical location for users. In this way, recommending products suitable for the user's temporal and spatial demands in LBS is a pretty challenging problem.
 



Recently, some methods\cite{din,dien,dsin} have been proved effective in e-commerce through the user's historical behavior, but it is not easy to adapt them into the LBS scenario. The main reason is that most of them do not pay attention to users' strong spatial and temporal demands. For instance, a user prefers fast food in the work area on weekdays and may choose fried chicken in his or her residential area on weekends. This changes in user behavioral interests are bonded with the changes of location and time. Although there are some initial efforts\cite{ST-PIL,LTSCR} to integrate spatiotemporal information into sequential recommendation, most of them consider partial spatiotemporal information, and efforts to fully and thoroughly model such integrated spatiotemporal patterns are still lacking. Different from the above scenarios, there are some common attributes in the takeaway scenario which have a weak correlation with the user's historical behavior. For example, milk tea is naturally suitable to be recommended at afternoon tea. On the other hand, the historical behaviors of users imply their personal dietary preferences.

To tackle above problems, we propose a \textbf{S}patio\textbf{t}emporal-\textbf{E}nhanced \textbf{N}etwork(\textbf{StEN}), to better meet users' temporal and spatial demands. Specially, StEN applies \textbf{S}patio\textbf{t}emporal \textbf{Pro}file Activation (\textbf{StPro}) module to model user's common spatiotemporal preference by activating attribute features (user and item). For the 
personalized spatiotemporal preference of users, a novel \textbf{S}patio\textbf{t}emporal \textbf{Pre}ference Activation (\textbf{StPre}) and  a \textbf{S}patio\textbf{t}emporal-aware \textbf{T}arget \textbf{A}ttention (\textbf{StTA}) module are proposed. StPre disassembles the spatiotemporal preference embodied by the user's historical behavior in detail, which including Temporal Evolving Activation(TEA), Temporal periodic Fusion(TPF) and Spatial Preference Activation(SPA). While StTA employs different spatiotemporal information to generate different parameters and feed them into target attention to improve the personalized spatiotemporal awareness of the model. In addition, we have released an industrial dataset for takeaway industry to make up for the lack of public datasets in this community.





All our contributions can be summarized as follows: 
\begin{itemize}
    \item StEN applies \textbf{S}patio\textbf{t}emporal \textbf{Pro}file Activation (\textbf{StPro}) module to model user’s common spatiotemporal preference by activating attribute features (user and item).
    \item For the personalized spatiotemporal preference of users, a novel \textbf{S}patio\textbf{t}emporal \textbf{Pre}ference Activation (\textbf{StPre}) is proposed, which disassembles the spatiotemporal preference embodied by the user's historical behavior in detail, and extracts preferences from three small modules: Temporal Evolving Activation (TEA), Temporal Periodic Fusion (TPF) and Spatial Preference Activation (SPA).
    \item We also propose a \textbf{S}patio\textbf{t}emporal-aware \textbf{T}arget \textbf{A}ttention (\textbf{StTA}) module, which employs different spatiotemporal information to generate different parameters and feed them into target attention to improve the personalized spatiotemporal awareness of the model
    \item In addition, we have also released an industrial dataset for takeaway industry to make up for the lack of public datasets in this community. Experimental results demonstrate that our method has achieved the state-of-the-art on three large-scale industrial datasets and the online A/B testing results further show its practical value.
\end{itemize}
\section{Related Work}
\subsection{Sequence-based Model}
Earlier deep CTR approaches hope to eliminate the complicated work of feature engineering jobs and focus more on automatically mining the correlations between features\cite{w&d, DeepFM, DCN, xDeepFM, autoInt}. Later on, researchers\cite{din, dien, dsin} found that the users' historical behavior sequence contains richer and more direct information, which brought breakthroughs to the entire recommendation community. Many researches focus on exploring potential interests in the user's historical behavior sequence. They extract sequence features by incorporating structures such as Pooling, RNN, and Attention into the model. YoutubeDNN\cite{google_youtube} proposes a feature embedding on items method and then takes the average value to extract historical sequence features. DIN\cite{din} believes that interests in the user's historical behavior sequence are diverse. 
Faced with a particular product, only part of the interests associated with that product will influence user's behavior. Based on this, DIN designs a local activation module to extract different user interests from the sequence for various target commodities. DIEN\cite{dien} further explores the interrelationships between users' historical behaviors and proposes the concept of user interest evolution. It designs an auxiliary loss and a structure based on GRU. Inspired by the success of the self-attention mechanism in sequence-to-sequence tasks, BST\cite{bst} leverages a transformer layer instead of GRU to mine information about the user's interest. DSIN\cite{dsin} observes that the user's interests in a short period are concentrated, while long-term interests are scattered. It splits the sequence into different sessions and explores the information through the self-attention mechanism and Bi-LSTM module. SIM\cite{SIM} proposes an interest mining method for life-long user sequences. However, all historical behavior sequences of users are very long, which may lead to time-consuming and noise problems. To overcome this, SIM provides a search-based long sequence extraction method to extract top-k behavior sequences from life-long sequences through soft and hard search technology.

\begin{figure*}[tp]
  \centering
  {
      \includegraphics[width=2\columnwidth]{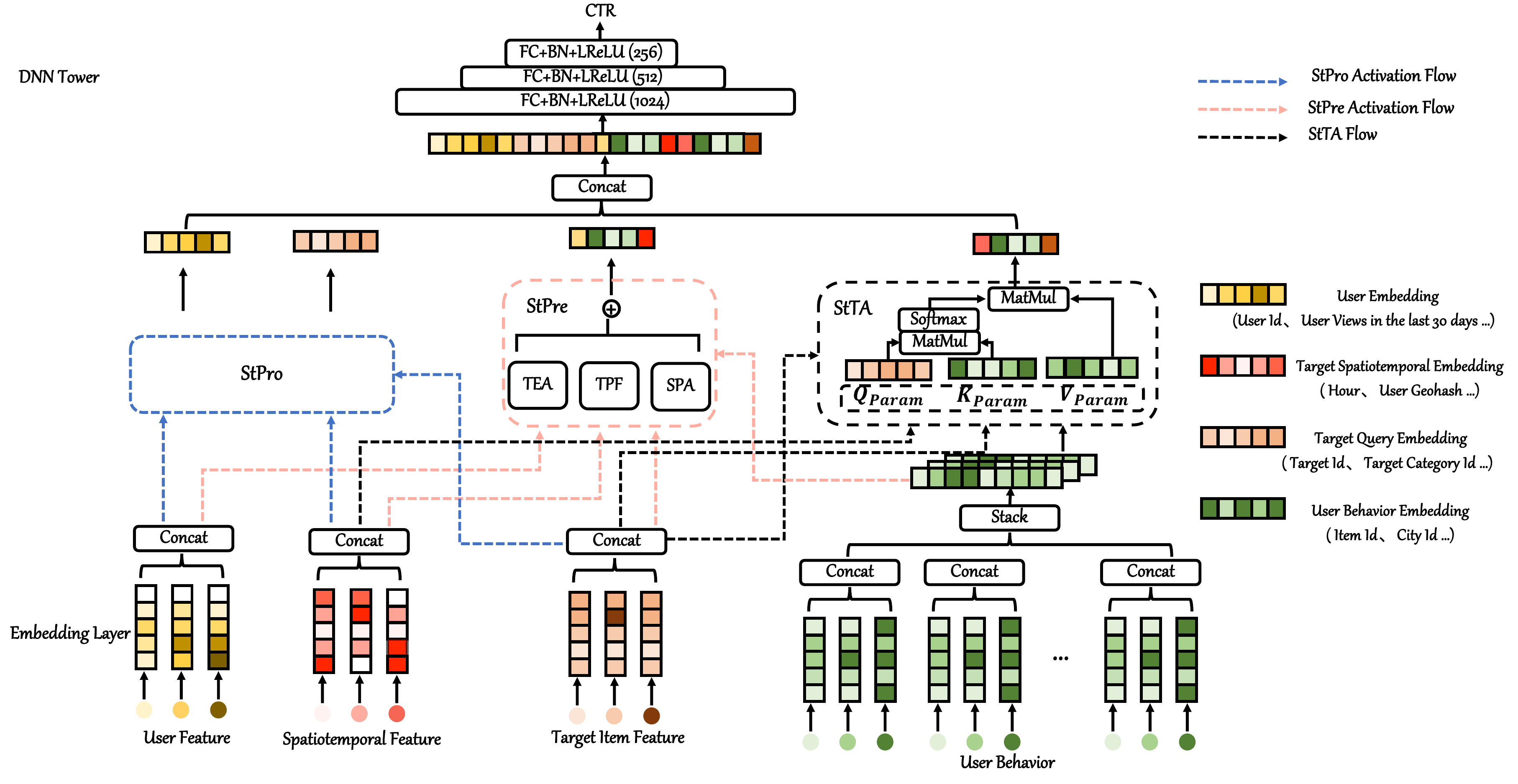}
  }
  \caption{Our StEN consists of three modules: Spatiotemporal Profile Activation(StPro), Spatiotemporal Preference Activation(StPre) and
Spatiotemporal-aware Target Attention(StTA).}
  \label{fig:framework}
\end{figure*}

\subsection{Time Aware Attention Model}
The above deep CTR models do not explicitly make use of the click time information in the user's historical behavior, where the click time information has an impact on the user's evolutionary behavior and the user's periodic behavior. The user's evolutionary behavior denotes that the user's interest changes over time, and the user's periodic behavior indicates the user's periodic actions. Specially, TIEN\cite{TIEN} pays more attention to the user's evolutionary behavior, and believes that the closer the historical behavior is to the current time, the greater the weight should be. TLSAN\cite{TLSAN} leverages the absolute value of the time difference and then uses its reciprocal as the time position embedding.  TiSASRec\cite{TiSASRec} models items' relative time intervals by sine and cosine function to explore the evolutionary behavior of users and then utilizes items' absolute temporal signals, such as month(M), weekday(W), date(D) and hour(H), to detect periodic behavior of users. TimelyRec\cite{TimelyRec} captures potential irregularity information in user's periodic patterns, and then integrates the information to compute the similarity between target time and users interactions with an attention mechanism.

\subsection{Spatiotemporal Model}
Spatial location is also important for some location-aware platforms, such as Facebook Places\cite{Facebook} and Airbnb\cite{Airbnb}. Thus, it is a natural way to integrate temporal information and spatial location to optimize recommendation models. However, due to the complexity of model design, publicly available existing work is limited. CaledarGNN\cite{CaledarGNN} utilizes GNN and GRU to extract the segmented time and geographic information in the user's historical behavior sequence. While effective, it is applied to article browsing of web pages without regard to the geographic location of the item. So it is not suitable for our takeaway industry. TRISAN\cite{TRISAN} extracts the spatiotemporal information from the user's historical behavior sequence by employing two spatial activation and one temporal similarity activation modules in the model. However, it does not detail the information contained in the user's spatiotemporal behavior, which leads to insufficient spatiotemporal information exploring. While TRISAN is of great relevance for our purposes, unfortunately, the method has not been open-sourced and the dataset used in this paper is not publicly available. So we cannot perform method comparisons with it in the Section~\ref{section:experiments}.


\section{Spatiotemporal-Enhanced Network}
\subsection{Preliminary}
\label{section:preliminaty}
In this paper, we denote $x=(m,u,st,b) \in \mathcal{X}$ as input data, where $m$ is the target item feature, $u$ is the user, $b$ is the user click behavior and  $st$ is the spatiotemporal feature. 

In particular, we geocode$\footnote{https://en.wikipedia.org/wiki/Geohash}$ the user's latitude and longitude and convert them to hexadecimal numbers to obtain geohash-6, which is then combined with the user's Area-of-Interest(AOI)\cite{aoi} and serve as the spatial feature $g$ in this paper. While the temporal feature is represented by hour of day, time period of day(breakfast, lunch, afternoon tea, dinner and night snack) and day of the week. User features $u$ include user id, user gender and other features, while item features $i$ include item id, item category and other features. Before all features enter the model, we will perform a vectorized representation of them. For the convenience of description, in the latter part of this article, $m,u,st,b$ all represent the embedding vectors of the corresponding features. Denoting $y \in \mathcal{Y}$ as the click label, and our CTR prediction task can be defined as:
\begin{equation}
  \mathcal{P}(y=1|x) = f(x; \theta)(x\in \mathcal{X})
  \label{eq-1}
\end{equation}
where $f(x; \theta)$ is a probability value obtained after we forward the input data $x$ into any CTR network, and then activate by a sigmoid function. $\theta$ represents the parameters of the network. Typically, each of our user history behaviors includes the item $v$, the item's location $l$, the click time $t$ and the click period of time $p$. The CTR task of Equation~\ref{eq-1} above is then mainly achieved by minimizing the following cross-entropy loss function during training,
\begin{equation}
  \mathcal{L}(f,x_i,y_i) = \frac{1}{N}\sum_{i=1}^{N}-y_ilogf(x_i; \theta)-(1-y_i)log(1-f(x_i; \theta))
  \label{eq-2}
\end{equation}
where $y_i\in \{0, 1\}$  is the ground-truth label, $N$ is the mini-batch size and $i$ is the index of the input data. We set $N$ to 1024 in this paper.



\subsection{Spatiotemporal Profile Activation}
 This module is mainly used to capture common spatiotemporal preferences that are less correlated with user behavior. E-commerce scenarios only need to consider the personalized behavior of user, but in the takeaway scenario, we need to consider the impact of time and location on users and items. For instance, there is a natural difference between the user's order in the workplace and the residential area. Therefore, we use spatiotemporal features $st$ to extract common spatiotemporal preference for the static item and user features. Below we will take the user feature as an example,
\begin{equation}
  att_{u} = sigmoid(\frac{FC_{u}(st) \cdot u^T}{\sqrt{d_{u}}}) u
  \label{eq-2.5}
\end{equation}
where $FC_{u}(st) \in \mathbb{R}^{d_{st} * d_u} $, is the linear transformation of the $st$, $d_{u}$ is the last dimension of $u$, $d_{st}$ is the last dimension of $st$. Inspired by \cite{din}, we then concatenate $u$ and $att_u$ and add their differences, their common values, to get the final activation value $h_{u}=concat(u, att_u, u-att_u, u*att_u)$.

\begin{figure}[tbp]
\centerline{\includegraphics[width=1.\columnwidth]{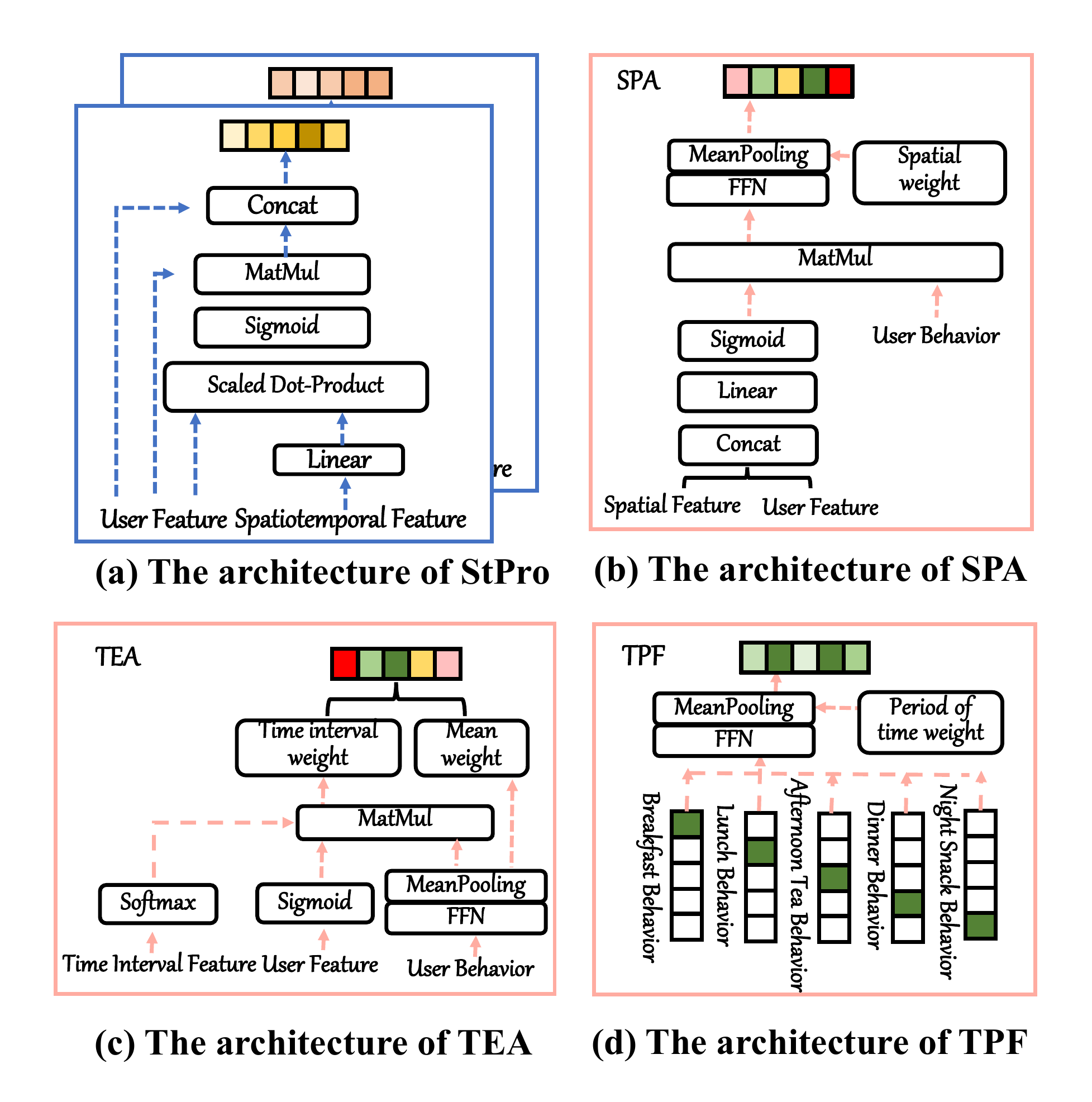}}
\caption{The architecture of Spatiotemporal Profile Activation(StPro) and Spatiotemporal Preference Activation(StPre). StPre includes three models: Temporal Evolving Activation(TEA), Temporal periodic Fusion(TPF) and Spatial Preference Activation(SPA).}
\label{fig:stpre_stpro}
\end{figure}

Through the above same activation method, we can obtain the final activation value of the item and is denoted as $h_{m}$. Finally, we concat the above activation values to obtain the spatiotemporal profile activation value $h_{stpro}$. Fig.~\ref{fig:stpre_stpro}(a) shows the structure.


\subsection{Spatiotemporal Preference Activation}
We further propose a Spatiotemporal Preference Activation(Stpre) to model the personalized spatiotemporal preference embodied by user behaviors in detail.
\subsubsection{Temporal Evolving Activation(TEA)}
The time sequence of user clicks will have a certain impact on the current behavior. For example, a user who frequently clicks on milk tea in a short period of time will cause him to be more willing to click on dessert in the next time slot. To model this temporal evolving pattern, we first calculate the time interval $t_{i}$ between request time $t_r$ and each historical behavior click time $t^j$. Then we eliminate the noise by applying a nonlinear transformation to the time interval, thus obtaining the temporal evolution factor $f_{te}$, 
\begin{equation}
  f_{te} = FC_2(LeakyReLU(FC_1(e^{-t_{i}}))) + e^{-t_{i}}
  \label{eq-3}
\end{equation}
where $FC_1 \in \mathbb{R}^{N * N_h} $ and $FC_2 \in \mathbb{R}^{N_h * N_l}$ denotes two fully connected layers, $t_{i} \in \mathbb{R}^{N * N_l}$, $N_h$ is the hidden size, and $N_l$ is the sequence length we set. In this paper, we abbreviate the structure of Equation~\ref{eq-3} as FFN. Then we normalize the above temporal evolution factor $f_{te}$ through a softmax function to get the weight of temporal evolution $w_{te}$. After that, $w_{te}$ can help to obtain temporal activation features related to the behavior order,

\begin{equation}
  att_{tea} = w_{te} \cdot FFN(sigmoid(FC_{t}(u)) \cdot b)
  \label{eq-6}
\end{equation}
where $FC_{t}(u) \in \mathbb{R}^{N_u * 1}$, $N_u$ is the last dimension of the feature $u$. Finally our robust temporal evolution fusion feature be obtained by $h_{tea} = w_{m} * MeanPooling(FFN(b)) + w_{tea} * att_{tea}$. Mean weight $w_{m}$ and time interval weight $w_{tea}$ are two trainable weight parameters used to balance the output. The module is depicted in Fig.~\ref{fig:stpre_stpro}(c).


\subsubsection{Temporal periodic Fusion(TPF)}
User historical behavior contains rich but scattered behavioral interests. However, when we explore user behavior from the perspective of time period, we are pleased to find that users' behavioral interests are more concentrated and periodic. Model would be messy if we directly learn mixed user behavior without any behavioral slices. In this case, we propose a Temporal periodic Fusion module to learn the user periodic preference in takeaway industry.


Based on the period of time $p$, we first divide the user historical behavior $b$ into five time slices $b=\{b_{pb},b_{pl},b_{pt},b_{pd},b_{ps}\}$. Then we feed each period of time sequence into the FFN and mean pooling in turn to get the characteristics of breakfast behaviors $mean_{pb}$, lunch behaviors $mean_{pl}$, afternoon tea behaviors $mean_{pt}$, dinner behaviors $mean_{pd}$, and night snack behaviors $mean_{ps}$. Take the breakfast behavior as an example,
\begin{equation}
  mean_{pb} = MeanPooling(FFN(b_{pb}))
  \label{eq-ffn_mp}
\end{equation}
Further, to obtain a more general periodic representation $h_{tpf}$, we fuse the above periodic characteristics through an average operation. Fig.~\ref{fig:stpre_stpro}(d) illustrates a outline of this architecture.


\subsubsection{Spatial Preference Activation(SPA)}
User's geographic location affects his personalized dietary choices. For example, when the user works in company, he may choose rice, and when the user is at home, he may prefer fried chicken. We call this the user's spatial preference. To capture this spatial preference, we utilize the spatial features $g$ and combine them with the user's feature $u$. We then feed the above-combined values into a fully connected layer and activate through a sigmoid function to get the geolocation activation value of $q_{spa}$,
\begin{equation}
  q_{spa} = sigmoid(FC_q(concat(g, u)))
\end{equation}
where $FC_q \in \mathbb{R}^{N_{gu} * 1}$, $N_{gu}$ is the dimension of the combine value $g$ and $u$. Further, we use $q_{spa}$ to activate all of the user history behavior to explore the user's spatial preferences $h_{spa}$ through FFN and mean pooling. The architecture can be observed in Fig.~\ref{fig:stpre_stpro}(b).

Finally, we fuse the output of the above three small modules together to obtain our final spatiotemporal preference activation value $h_{stpre} = h_{tef}+w_{tpf}*h_{tpf}+w_{spa}*h_{spa}$. Period of time weight $w_{tpf}$ and spatial weight $w_{spa}$ are also two trainable weight parameters used to balance the output.

\subsection{Spatiotemporal-aware Target Attention}
\label{section:stta}
To more effectively explore the spatiotemporal relationships between historical user behavior and target item, we propose a Spatio-temporal-aware Target Attention(StTA) mechanism. Drawing on the ideas of CAN\cite{CAN} and AdaptPGM\cite{AdaptPGM}, we generate different parameters through spatiotemporal information for target attention, thereby improving the personalized spatiotemporal awareness of the model. Taking ${W_Q,b_Q}$ as an example, we can get that,
\begin{equation}
    Q_{Param}=W_q\cdot{st}+b_q\to{W_Q,b_Q}
\end{equation}
where $W_q\in{\mathbb{R}^{D\times{(d_i*d_o +d_o)}}}$ and $b_q\in{\mathbb{R}^{d_i*d_o+d_o}}$ are the parameters of a fully-connected layer. $D$ is the dimension of $st$, $d_i$ is the dimension of input embedding (such as target item embedding $m$ or user behavior embedding $b$) and $d_o$ is the dimension of final output embedding. Then we can split $Q_{Param}$ into two parts($W_Q$, $b_Q$) as parameters of the subsequent target attention fully connected layer. Specially, we take the first $d_i*d_o$ parameters as $W_Q$ and the last $d_o$ parameters as $b_Q$. In the same way, we can obtain $K_{Param}$($W_K$, $b_K$) and $V_{Param}$($W_V$, $b_V$) through the spatiotemporal feature $st$. After that, we utilize the primitive target attention mechanism to obtain the final module output $h_{ta}$,
\begin{gather}
    Q=W_Q\cdot{m}+b_Q ,
    K=W_K\cdot{b}+b_K ,
    V=W_V\cdot{b}+b_V  \\
    h_{ta}=softmax(\frac{QK^T}{\sqrt{d_K}})V
\end{gather}
Where $d_k$ is the dimension of $K$. Fig.~\ref{fig:framework}(a) illustrates the structure.


\subsection{Dense Tower for StEN}

Once we have all the feature vector representations, we can fuse all the above module outputs to get the final prediction $dense_0= concat(h_{stpro}, h_{stpre}, h_{ta})$. A three-layer perceptron structure is then applied,
\begin{equation}
  dense_{i+1} = LeakyReLU(BatchNorm(FC_{fi}(dense_i)))
\end{equation}
where $i=0,1,2$. We then get the prediction of click via a sigmoid activation $\mathcal{P}(y=1|x) = sigmoid(FC_{sigmoid}(dense_{3}))$. $FC_{sigmoid} \in \mathbb{R}^{N * 1} $. Finally, we optimize the parameters of our whole model by Equation~\ref{eq-2} defined above. The detail is illustrated in Fig.~\ref{fig:framework}(a).


\begin{table}
  \caption{Statistics of the dataset used in this paper. ML indicates median length.}
  \input{./table/dataset.tex}
  \label{table:intr_dataset}
\end{table}

\section{EXPERIMENTS}
\label{section:experiments}
\subsection{Datasets}
Due to the lack of public spatiotemporal datasets in the takeaway industry, we conducted experimental comparisons on three industrial datasets ($D_1$, $D_2$ and $D_3$) collected from Ele.me, a major LBS platform in China. The dataset $D_1$ mainly recommend stores to users, which consists of over 5 billion samples. Dataset $D_2$ and $D_3$ mainly recommend meals to users and contain more than 500 million and 100 million samples, respectively. For $D_3$, we collected one week's data from the server logs as training set and one day's data as the test set. We have publicly released the dataset $D_3$ $\footnote{https://tianchi.aliyun.com/dataset/dataDetail?dataId=131047}$ to further advance the exploration of spatiotemporal patterns in the LBS community. The details of 
our datasets can be seen in Table~\ref{table:intr_dataset}.

\begin{table}
  \begin{center}
      \caption{Overall performance on $\mathcal{D}_{1}$, $\mathcal{D}_{2}$ and $\mathcal{D}_{3}$. StPro: Spatiotemporal Profile Activation. StPre: Spatiotemporal Preference Activation. DIN+StPro+StPre, DHAN+StPro+StPre, DIEN+StPro+StPre are three variation models to investigate the generalization of our module.}
      \input{./table/sota.tex}
      \label{table:pfm_sota}
  \end{center}
\end{table}

\subsection{Experimental Settings}
All models in this paper are implemented with Pyhton 2.7 and Tensorflow 1.4. AdagradDecay\cite{adgraydecay} is chosen as our optimizer to train the model. To avoid overfitting in the early stage of model training and maintain the training stability, we adopt a warm-up\cite{warmup} strategy for all methods. We set the learning rate to 0.001 and gradually increased it to 0.015 within 1M steps. We set the batchsize $N$ to 1024. We repeated all the experiments five times and averaged the metrics to obtain more reliable results. In our experiments, We adapt Area Under Cure (AUC) and RelaImpr\cite{relaimpr} as our evaluation metric.

To show the effectiveness of our method, we select three well-known and industry-proven CTR prediction models as our baselines.

\textbf{DIN}: Deep Interest Network (DIN) designs a local activation module to capture the information in the user behavior sequence that will affect the user behavior when facing the target item. At the same time, DIN does not model the interrelationships among items in a sequence of actions.

\textbf{DHAN}: Deep Hierarchical Attention Networks(DHAN) designs a set of attention networks with multi-dimensional and multi-level structures, which can capture the interest expression of users in various dimensions. At the same time, the attention network can extract features that are similar to the knowledge expression of the tree structure.

\textbf{DIEN}: Deep Interest Evolution Network (DIEN) adapts the interest evolution factors in user behavior. It designs an AUGRU-based module to model the evolution process and trend of user interests.

\begin{figure*}[tbp]
    \centering
    \subfigure{
        \includegraphics[width=0.8\columnwidth]{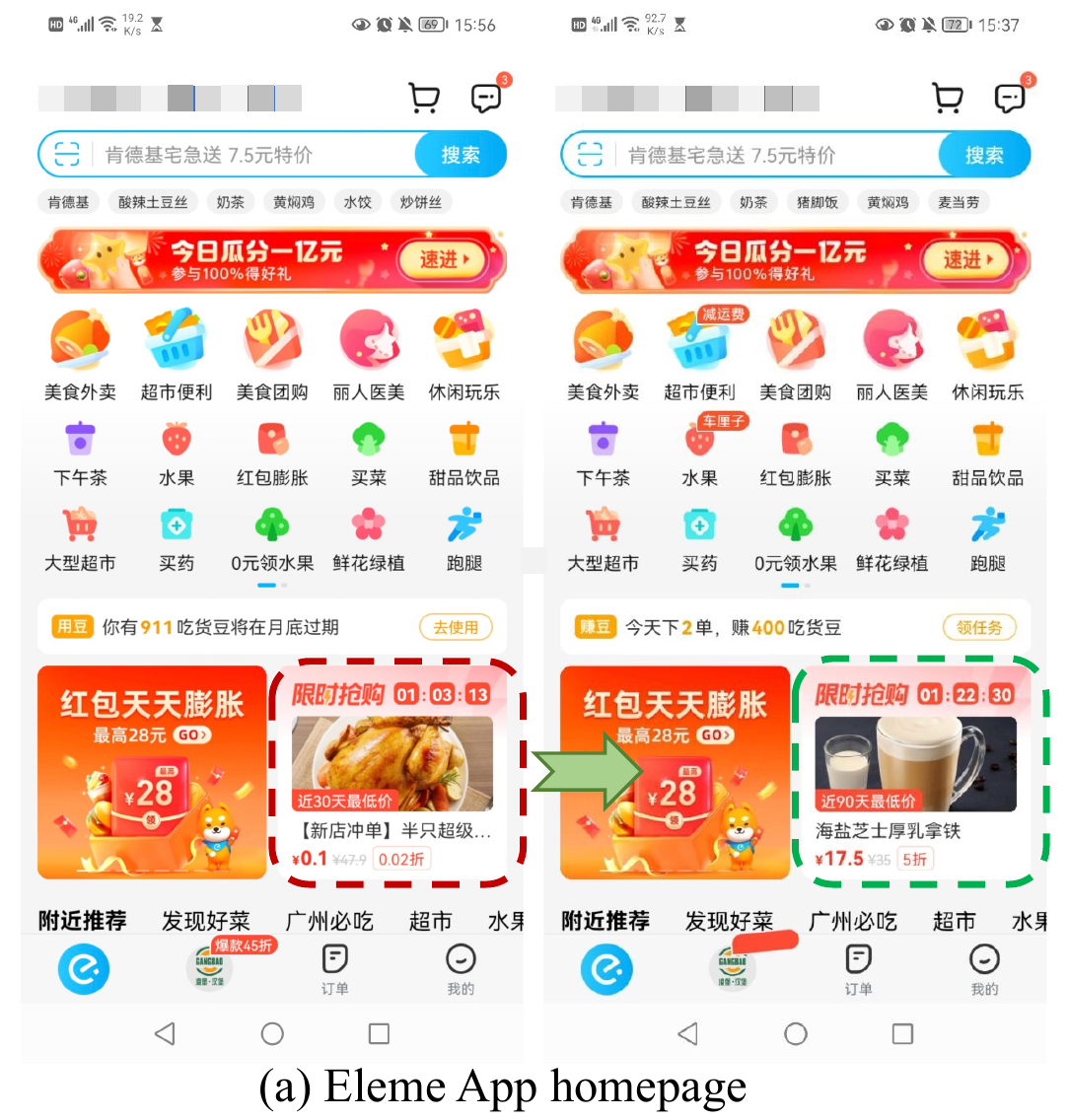}
        \label{fig:case1}
    }
    \subfigure{
        \includegraphics[width=0.8\columnwidth]{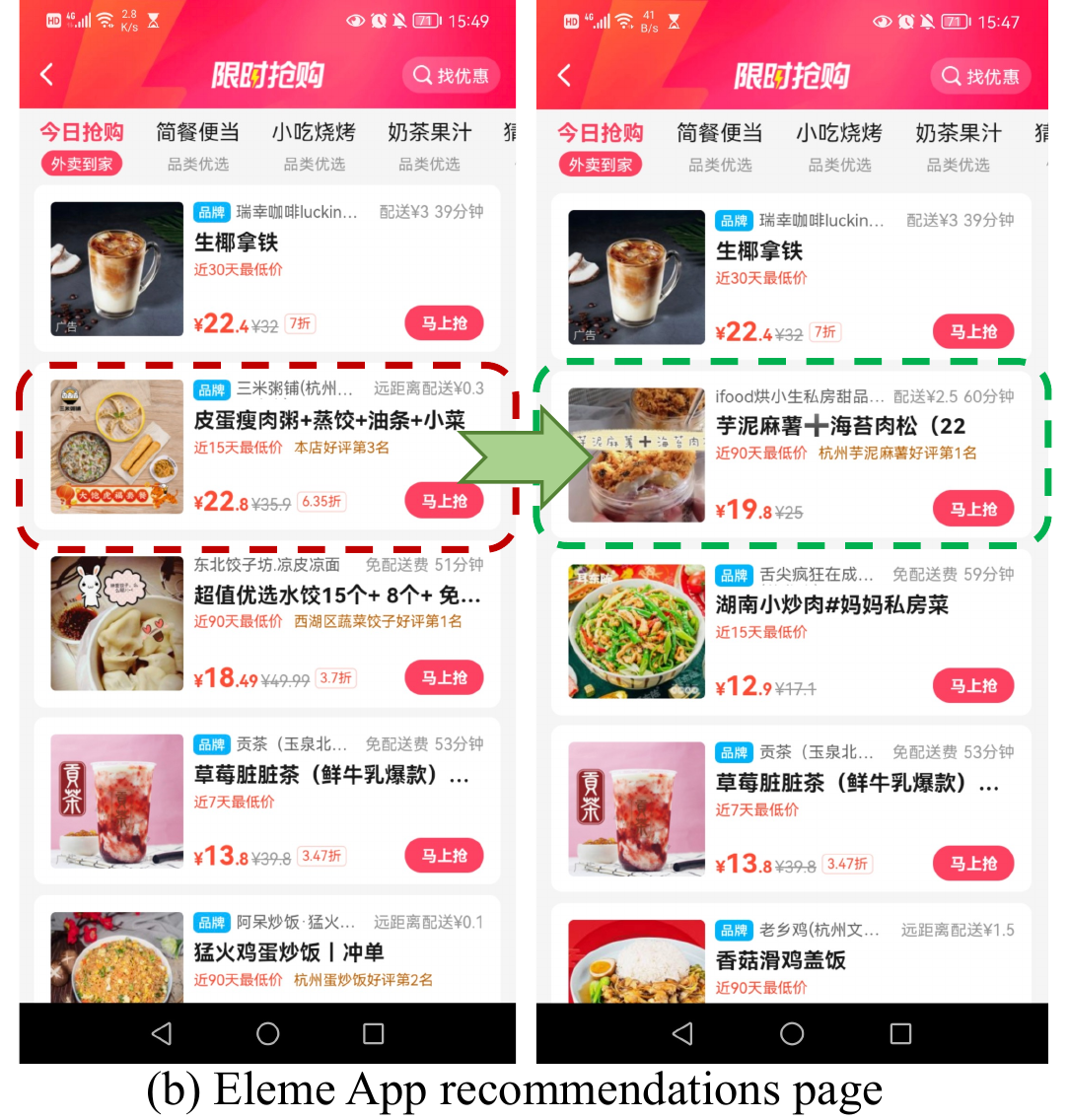}
        \label{fig:case2}
    }
    \caption{Screenshots of the Eleme mobile App. (a) and (b) are the recommendation results of the online-serving model (red box) and StEN (green box) during afternoon tea, where the right of (a) and (b) (green box) are more suitable for afternoon tea.}
    \label{fig:case_study}
\end{figure*}

\subsection{Overall Performance Comparison}
Table~\ref{table:pfm_sota} compares StEN with three well-known CTR prediction models on $D_1$, $D_2$ and $D_3$.  We find that DHAN\cite{dhan} performs better than DIN\cite{din} on both datasets due to the addition of a multi-dimensional and multi-level attention mechanism. For example, DHAN surpass DIN on by margins of 0.56\% on dataset $D_1$. Notably, 0.1\% improvement of AUC is significant for online model deployment to improve the actual CTR in production. Due to the excellent performance of LSTM module in exploring user behavior sequence, DIEN\cite{dien} outperforms DHAN in both datasets. However, it is worth noting that recurrent neural networks such as LSTM have slow training and prediction problems and are prone to high response time problems when serving online. By comparison, our StEN advantages all of them to a new level. We have achieved \textbf{AUC=0.7353}, \textbf{AUC=0.7525} and \textbf{AUC=0.6627} on $D_1$, $D_2$ and $D_3$, respectively. Our method is 0.96\% higher than current best results (DIEN) on dataset $D_3$. 

At the same time, to investigate the generalization of our module, we have conducted variation experiments by adding StPre and StPro to the above baseline models. Note that the main difference among the above three methods is the attention module, so our StTA will not be added to interfere. It can be observed from Table~\ref{table:pfm_sota} that when we directly adapt our two proposed activation modules to the three baselines mentioned above, there is a certain stable improvement in performance. For example, DIN obtains a significant improvement of 0.27\% on $D_1$ , 0.30\% on the $D_2$  and 0.31\% on the $D_3$, while DIEN has the weaker improvement of 0.02\% on $D_1$ , 0.06\% on $D_2$ and 0.4\% on the $D_3$.  All these variation models further demonstrate that our proposed modules have good generalizability and can be added to other existing models as a plug-and-play module.



\begin{table}
  \begin{center}
      \caption{Ablation study on $\mathcal{D}_{1}$ and $\mathcal{D}_{2}$. StPro: Spatiotemporal Profile Activation. TEA: Temporal Evolving Activation. TPF: Temporal Periodic Fusion. SPA: Spatial Preference Activation. StPre: Spatiotemporal Preference Activation. StTA: Spatiotemporal-aware Target Attention.}
      \input{./table/ablation.tex}

      \label{table:ablation}
  \end{center}
\end{table}

\subsection{Ablation Study}
\label{sec-abl}
 To investigate the effectiveness of our proposed method, we conduct ablation studies in Table~\ref{table:ablation}. Our BaseModel in this paper consists of a primitive Target Attention module mentioned in Section~\ref{section:stta}. Observed from Table~\ref{table:ablation}, each module has played a different positive role after being added.

We then show the effect of Spatiotemporal Profile Activation (StPro) by adding it to the BaseModel. Observed From Table~\ref{table:ablation}, we can see that our "w/ StPro" has brought a relatively stable improvement in effect. In particular, compared to BaseModel, the offline AUC rises from 0.7332 to 0.7345 (+0.13\%)  and 0.7414 to 0.7474 (+0.6\%) when tested on $D_1$ and $D_2$, respectively. The results demonstrate that Spatiotemporal Profile Activation is an effective way to model user’s common spatiotemporal preference.

Next, we validate the effectiveness of Spatiotemporal Preference Activation (StPre) over the model. As reported in table~\ref{table:ablation}, "w/ StPre" increases the results of "BaseModel" by 0.17\% and by 1.07\% on dataset of $D_1$ and $D_2$, respectively. In order to see the effect of the three small modules (TEA, TPF and SPA) in StPre, we also performed some ablation experiments in Table~\ref{table:ablation}. We can observe that module SPA shows the best performance when tested on dataset $D_1$, while module TEA achieves better performance when tested on dataset $D_2$. This illustrates that in different scenarios, the user's spatiotemporal preferences will focus on different emphasis, specific focus needs to be specifically determined.

We also evaluate the effect of Spatiotemporal-aware Target Attention (StTA) mechanism. In Table~\ref{table:ablation}, we observe a significant improvement after adding Spatiotemporal-aware Target Attention into the system. For example, "w/ StTA" achieves an offline AUC of 0.7350 when tested on the dataset of $D_1$. This is higher than "BaseModel" by 0.18\%. The improvement demonstrates that our proposed Target Attention mechanism can meet the user's spatiotemporal demands compared to the primitive target attention module. Injecting our StTA into the model could improve the effectiveness of system in LBS. Furthermore, our "StEN(StPre+StPro+StTA)" consistently improves the results of "w/ StPre", "w/ StPro" and "w/ StTA". This is because more appropriate spatiotemporal enhancement has been conducted by integrating the three module we proposed in this paper. 
 
 
 


\subsection{Online A/B Testing}
We have deployed our method on the Ele.me platform and conducted an online A/B test for one month in November 2021, which is under the bucket test. One bucket is the BaseModel we have defined in Section~\ref{sec-abl} and the other bucket is our model StEN. Compared with the online-serving BaseModel, our method has increased the CTR of one-hop by 1.6\%, the CTR of the second-hop by 2.4\%, the order volume by 2.1\%, and the order UV by 2.4\%. These online benefits from our method are crucial for the recommendation systems of Ele.me.
On the one hand, an efficient model can improve user click efficiency. On the other hand, the emphasis on spatiotemporal characteristics can also improve user experience and increase the user stickiness of the platform.

For better understanding, we also compare the recommendation results of the online-serving model with our StEN on the Ele.me platform, as shown in Figure~\ref{fig:case_study}. The items (red box) on the left of Figure~\ref{fig:case_study}(a) and Figure~\ref{fig:case_study}(b) are not suitable for afternoon tea, but are more appropriate for breakfast and staple food, respectively. While our StEN (green box) recommends the sweetmeats and milk tea that are suitable for afternoon tea. Therefore, StEN does a better job of capturing users' strong spatial and temporal demands and can improve the user experience.

\section{Conclusions}
In this paper, we propose a novel spatiotemporal-enhanced network StEN. In particular, StEN applies a StPro module to capture common spatiotemporal preference by activating attribute features. A StPre module is further applied to model the personalized spatiotemporal preference embodied by the behaviors in detail. Moreover, a StTA mechanism is adopted to generate different parameters for target attention at different locations and times, thereby improving the personalized spatiotemporal awareness of the model. Comprehensive experiments are conducted on three large-scale industrial datasets, and the results demonstrate the state-of-the-art performance of our methods.


\bibliographystyle{ACM-Reference-Format}
\bibliography{./bib/paper}

\end{document}

%% file: table/dataset.tex

\begin{tabular}{cccc}
    \hline
    Datasets & $\mathcal{D}_{1}$ & $\mathcal{D}_{2}$ & $\mathcal{D}_{3}$ \\
    \hline\hline
    Total Size  & 5541799773 &  575941170 & 177114244 \\
    \# Feature & 388 & 218 & 38  \\
    \# Users  & 49249999 & 28706270 & 14427689  \\
    \# Items  & 2750505 & 12302502 & 7446116  \\
    \# Clicks  & 343277081 & 5626279 & 3140831  \\
    ML of User Behaviors   & 39.66 & 41.59 & 41.19 \\
    \hline
\end{tabular}

%% file: table/sota.tex
    

\begin{tabular}{c|c|c|c}
    \hline
    Model & $\mathcal{D}_{1}$ & $\mathcal{D}_{2}$ & $\mathcal{D}_{3}$ \\
    \hline\hline
    DIN & 0.7209 & 0.7294  & 0.6403\\
    \hline
    DHAN & 0.7265 & 0.7312 & 0.6419  \\
    \hline
    DIEN & 0.7346 & 0.7452 & 0.6531 \\
    \hline\hline
    DIN+StPro+StPre & 0.7236 & 0.7324 & 0.6434\\
    \hline
    DHAN+StPro+StPre  & 0.7271 & 0.7336  & 0.6445 \\
    \hline
    DIEN+StPro+StPre & 0.7348 & 0.7458 & 0.6571 \\
    \hline\hline
    \textbf{StEN} & \textbf{0.7353} & \textbf{0.7535} & \textbf{0.6627} \\
    
    \hline
\end{tabular}

%% file: table/ablation.tex
\begin{tabular}{c|cccc}
    \hline
    \multirow{2}*{Methods} & \multicolumn{2}{c}{$\mathcal{D}_{1}$} & \multicolumn{2}{c}{$\mathcal{D}_{2}$} \\  \cline{2-5}
    & AUC & RelaImpr & AUC & RelaImpr \\
    \hline
    BaseModel & 0.7332 & 0.00\% & 0.7414 & 0.00\% \\
     w/ StPro & 0.7345 & 0.56\% & 0.7474 & 2.49\% \\
     w/ TEA & 0.7345  & 0.56\% & 0.7500  & 3.56\% \\
     w/ TPF & 0.7342 & 0.43\% & 0.7479  & 2.69\% \\
     w/ SPA & 0.7348 & 0.69\% & 0.7476 & 2.57\% \\
     w/ StPre & 0.7349 & 0.73\% & 0.7521 & 4.43\% \\
     w/ StTA & 0.7350 & 0.77\% & 0.7499 & 3.52\% \\
    \hline\hline
    \textbf{StEN} & \textbf{0.7353}  & \textbf{0.90\%} & \textbf{0.7535} & \textbf{5.01\%} \\
    
    \hline
\end{tabular}